# Probe of Spin Dynamics in Superconducting NbN Thin Films via Spin Pumping

Yunyan Yao[1,2], Qi Song[1,2], Yota Takamura[3,4], Juan Pedro Cascales[3], Wei Yuan[1,2], Yang Ma[1,2], Yu Yun[1,2], X. C. Xie[1,2], Jagadeesh S. Moodera[3,5], and Wei Han[1,2*]

[1]International Center for Quantum Materials, School of Physics, Peking University, Beijing 100871, China.

[2]Collaborative Innovation Center of Quantum Matter, Beijing 100871, China.

[3]Plasma Science and Fusion Center and Francis Bitter Magnet Laboratory, Massachusetts Institute of Technology, Cambridge, MA 02139, USA.

[4]School of Engineering, Tokyo Institute of Technology, Tokyo 152-8550, Japan.

[5]Department of Physics, Massachusetts Institute of Technology, Cambridge, MA 02139, USA

*Correspondence to: weihan@pku.edu.cn (W.H.)

**Abstract:**

The emerging field of superconductor (SC) spintronics has attracted intensive attentions recently. Many fantastic spin dependent properties in SCs have been discovered, including large magnetoresistance, long spin lifetimes and the giant spin Hall effect, etc. Regarding the spin dynamics in superconducting thin films, few studies has been reported yet. Here, we report the investigation of the spin dynamics in a *s*-wave superconducting NbN film via spin pumping from an adjacent insulating ferromagnet GdN film. A profound coherence peak of the Gilbert damping of GdN is observed slightly below the superconducting critical temperature of the NbN, which agrees well with recent theoretical prediction for *s*-wave SCs in the presence of impurity spin-orbit scattering. This observation is also a manifestation of the dynamic spin injection into superconducting NbN thin film. Our results demonstrate that spin pumping could be used to probe the dynamic spin susceptibility of superconducting thin films, thus pave the way for future investigation of spin dynamics of interfacial and two dimensional crystalline SCs.

# I. INTRODUCTION

The interplay between superconductivity and spintronics has been intensively investigated in the last decade [1-3]. In the content of superconductivity, the ferromagnetic magnetization has been found to play an important role in the superconducting critical temperature ($T_C$) in the ferromagnet (FM)/superconductor (SC) junctions [4-9]. Besides, unusual spin-polarized supercurrent has been observed in ferromagnetic Josephson junctions [2,3,10-13]. Furthermore, non-abelian Majorana fermions have been proposed for $p_x + ip_y$ superconductor surfaces driven by magnetic proximity effect, which has the potential for topological quantum computation [14]. In the content of spintronics [15,16], the cooper pairs in SCs have exhibited many fantastic spin-dependent properties that are promising for future information technologies. For instance, large magnetoresistance has been achieved in the spin valve with SCs between two ferromagnetic films [7,17], extremely long spin lifetimes and large spin Hall effect have been discovered for the quasiparticles of SCs, which exceeds several orders compared to the normal states above $T_C$ [18,19]. Despite these intensive studies of the spin-dependent properties in the FM/SC junctions, the spin dynamics of the superconducting thin films has not been reported yet.

Very intriguingly, it has been proposed recently to investigate the spin dynamics in superconducting films via spin pumping [20], a well-established technique to perform dynamic spin injection and to probe the dynamic spin susceptibility in various materials, including metal, semiconductors, Rashba 2DEGs, and topological insulators, etc [21-30].

In this letter, we report the experimental investigation of the spin dynamics in *s*-wave superconducting NbN thin films via spin pumping. A profound coherence peak of the Gilbert damping of GdN is observed slightly below the superconducting critical temperature ($T_C$) of the NbN in the NbN/GdN/NbN trilayer samples, which indicates dynamic spin injection into the NbN thin film. Besides, the interface-enhanced Gilbert damping probes dynamic spin susceptibility in the NbN thin films in the presence of impurity spin-orbit scattering, which is consistent with the recent theoretical study [20]. Our experimental results further demonstrate that spin pumping could be a powerful tool to study the spin dynamics in the emerging two dimensional SCs [31-33].

## II. EXPERIMENTAL DETAILS

The NbN (t)/GdN (d)/NbN (t) trilayer samples are grown on $Al_2O_3$ (~5 nm)-buffered thermally oxidized Si substrates by d.c. reactive magnetron sputtering at 300 °C in an ultrahigh vacuum chamber. The NbN layers are deposited from a pure Nb target (99.95%) in Ar and $N_2$ gas mixture at a pressure of 2.3 mTorr (20% $N_2$), and GdN films are deposited from a pure Gd target (99.9%) in Ar and $N_2$ gas mixture at a pressure of 2.8 mTorr (6% $N_2$). The NbN and GdN layers are of textured crystalline quality with a preferred direction along (111)-orientation, as evidenced by X-ray diffraction results (Fig. S1) [34]. After the growth, a thin $Al_2O_3$ layer (~ 10 nm) was deposited *in situ* as a capping layer to avoid sample degradation with air exposure. The Curie temperature of the GdN film was determined via the offset-magnetization as a function of the temperature using a Magnetic Properties Measurement System (MPMS; Quantum Design). The $T_C$ of the SC NbN thin films was measured by four-probe resistance technique as a function of the temperature in a Physical Properties Measurement System (PPMS; Quantum Design) using standard ac lock-in technique at low frequency of 7 Hz. The FMR spectra of the multilayer samples were measured using the coplanar wave guide technique with a vector network analyzer (VNA, Agilent E5071C) in the variable temperature insert of PPMS. The samples were attached to the coplanar wave guide using insulating silicon paste. During the measurement, the amplitudes of forward complex transmission coefficients ($S_{21}$) were recorded as a function of the in-plane magnetic field from ~ 4000 to 0 Oe at various temperatures under different microwave frequencies and microwave power of 5 dBm.

## III. RESULTS and DISCUSSION

Fig. 1(a) illustrates spin pumping and the interfacial *s-d* exchange coupling between spins in the NbN layer and magnetic moments in the GdN layer. Due to the interfacial *s-d* exchange interaction ($J_{sd}$), the time-dependent magnetization in the GdN layer pumps a quasiparticles-mediated spin current into the NbN layer, and the spin-flip scattering of quasiparticles accompanies a quantum process of magnon annihilation in the GdN layer, giving rise to enhanced Gilbert damping [20]. Hence, the interfacial s-d exchange coupling provides the route to detect the spin dynamics of the NbN layer by measuring the magnetization dynamics of GdN. The spin pumping is performed by measuring Gilbert damping of the GdN film in the NbN (t)/GdN (d)/NbN (t) trilayer heterostructures via ferromagnetic resonance (FMR) technique [35]. NbN is a *s*-wave SC with short coherence length of ~ 5 nm and spin diffusion length of ~ 7 nm [19,36]. GdN is an insulating FM. A 10 nm NbN film is used for the NbN (t)/GdN (d)/NbN (t) samples to justify the assumption that the spin backflow from SC is small in the theoretical study [20]. In a typical sample of NbN (10)/GdN (5)/NbN (10) (with thickness in nm), the Curie temperature ($T_{Curie}$) of GdN is determined to be ~ 38 K from the temperature dependence of magnetic moments (Fig. 1(b)), and $T_C$ of NbN is obtained to be ~ 10.8 K via resistivity vs. temperature measurement (Fig. 1(c)). The $T_C$ is slightly affected by the external in-plane magnetic field of 4000 Oe due to the large critical field.

Fig. 1(d) shows a typical FMR signal ($S_{21}$) vs. the magnetic field measured at $T = 10$ K, with a microwave excitation frequency ($f$) of 15 GHz. The half linewidth ($\Delta H$) could be obtained by the Lorentz fitting of the FMR signal following the relationship [37]:

$$S_{21} \propto S_0 \frac{(\Delta H)^2}{(\Delta H)^2 + (H - H_{res})^2} \tag{1}$$

where $S_0$ is the coefficient for the transmitted microwave power, $H$ is the external magnetic field, and $H_{res}$ is the resonance magnetic field. Gilbert damping ($\alpha$) is determined using numerical fitting of $\Delta H$ vs. $f$ (Fig. 1(e)) based on the spin-relaxation mechanism (Fig. S2) [34,38]:

$$\Delta H = \Delta H_0 + \frac{4\pi\alpha f}{\gamma} + A\frac{2\pi f\tau}{1+(2\pi f\tau)^2} \tag{2}$$

where $\Delta H_0$ is related to the inhomogeneous properties, $\gamma$ is the gyromagnetic ratio, $A$ is the spin-relaxation coefficient, and $\tau$ is the spin-relaxation time constant.

The Gilbert damping of GdN is studied as a function of the temperature for two trilayer samples with the same interface: NbN (2)/GdN (5)/NbN (2) and NbN (10)/GdN (5)/NbN (10). The 2 nm nm NbN layer is not superconducting down to 2 K while the NbN (10)/GdN (5)/NbN (10) exhibits a $T_C$ of ~ 10.8 K (Fig. 2(a)). The non-superconducting feature of 2 nm NbN sample and lower $T_C$ of the 10 nm NbN sample compared to over 16 K for singly crystalline NbN could be attributed to reduced thickness, polycrystalline property, and magnetic proximity effect [1,36,39]. Interestingly, a profound coherence peak of the Gilbert damping is observed in the NbN (10)/GdN (5)/NbN (10), but not on NbN (2)/GdN (5)/NbN (2), as shown in Fig. 2(b). This feature is also evident from the temperature dependence of the half linewidth (Fig. S3) [34]. The peak of the Gilbert damping in the NbN (10)/GdN (5)/NbN (10) is observed at ~ 8.5 K, which is slightly below the $T_C$ (~ 10.8 K) of the 10 nm NbN layers. These results indicate the successful dynamic spin injection into the 10 nm NbN superconducting layer, which authenticates a charge-free method to inject spin-polarized carriers into SCs beyond previous reports of electrical spin injection [18,40-42]. Furthermore, the observation of the profound coherence peak at $T = \sim 0.8$ T$_C$ is expected based on recent theoretical studies of the spin dynamics for *s*-wave superconducting thin films via spin pumping [20]. Since the thickness of NbN layer (d = 10 nm) is longer than its spin diffusion length of $\sim$ 7 nm, the spin backflow effect, which is expected to reduce the damping peak, is negligible here [20,21]. According to the theory, Gilbert damping is related to the interfacial *s-d* exchange interaction and the imaginary part of the dynamic spin susceptibility of the SC [20].

$$\delta\alpha \propto J_{sd}^2 \sum_q \mathrm{Im}\, x_q^R(\varpi) \qquad (3)$$

For the *s*-wave superconducting NbN thin films, the superconducting gap $\Delta$ forms below T$_C$. At the temperature slightly below T$_C$, two coherence peaks of the density states exist around the edge of the superconducting gap following the BCS theory [43], and these peaks in turn give rise to the enhancement of the dynamic spin susceptibility in the presence of impurity spin-orbit scattering. Quantitatively, the ratio of the peak value of Gilbert damping over that at $T \sim$ T$_C$ is $\sim$ 1.8, which also agrees well with the theoretical calculation using previous experimental values of spin diffusion length ($\sim$ 7 nm), phase coherence length ($\sim$ 5 nm), and mean free path ($\sim$ 0.3 nm) for NbN thin films [19,20,36]. As the temperature further decreases, the number of quasiparticles decreases rapidly as the $\Delta$ grows, giving rise to the fast decrease of Gilbert damping below $T = \sim$ 7 K.

Next, the thickness of the GdN layer is varied to further study spin pumping and the spin dynamics in NbN layer. For both samples of NbN (10)/GdN (d)/NbN (10) with d = 10 nm and d = 30 nm, a profound coherence peak of Gilbert damping is observed slightly below the superconducting temperature of NbN, as shown in Figs. 3(a) and 3(b) (red circles). While for all the samples of NbN (2)/GdN (d)/NbN (2), no such coherence peak of the Gilbert damping is noticeable (green circles in Figs. 3(a) and 3(b)). The role of the effective magnetization in the observed coherence peak has been ruled out since it exhibits similar temperature dependence for NbN (2)/GdN (d)/NbN (2) and NbN (10)/GdN (d)/NbN (10) samples (Fig. S4) [34]. Clearly, a more profound damping peak for the d = 10 nm sample is observed compared to d = 5 nm and d = 30 nm samples. Fig. 3(c) shows the ratio of the peak Gilbert damping over the value at $T = \sim$ T$_C$ as a function of the GdN thickness, which is in the range from 1.8 to 2.8. The T$_C$ exhibit little variation as a function of the GdN thickness (Fig. S5) [34]. For a deeper understanding of the underlying mechanism to account for the thickness dependence of the ratio, further theoretical and experimental studies would be essential. One possible cause might be related to the interface proximity exchange effect and/or the presence of magnetic loose spins leading to scattering at the interface, which could affect the spin diffusion length, coherence length, and mean free path of the NbN layer.

To further confirm that the observed coherence peak of the Gilbert damping arises from the *s-d* exchange interaction at the SC/FM interface, the interface-induced Gilbert damping ($\alpha_S$) is studied as a function of temperature. $\alpha_S$ is obtained from the thickness dependence of total Gilbert damping at each temperature [34], as shown in Fig. 4(a). The contribution from GdN itself is significantly small compared to $\alpha_S$. The unperturbed peak at ~ 8.5 K (Fig. 4(b)) unambiguously demonstrates that the origin of the coherence peak in the Gilbert damping is indeed due to the interfacial *s-d* exchange interaction between the magnetization of GdN and the spins of the quasiparticles in superconducting NbN thin films. The slightly upturn of the interface-induced Gilbert damping starting at $T = 11$ K might be associated with the fluctuation superconductivity and/or the higher superconducting transition temperature for some regions of the NbN films than the zero resistance temperature. For comparison, the interface-induced Gilbert damping in the samples of NbN (2)/GdN (d)/NbN (2) does not exhibit any signature of coherence peak between 4 and 15 K (Fig. S6) [34].

Noteworthy is that our results are essentially different from previous reports of spin pumping into superconducting Nb films using the ferromagnetic metal permalloy (Py), where a monotonic decrease of the Gilbert damping is reported when the temperature decreases [44,45]. This feature is also confirmed in our studies by measuring the temperature dependence of Gilbert damping of Py in Nb(100)/Py(20)/Nb(100) (Fig. S7) [34]. Since Py is a FM metal, the interface exchange interaction would be significant to strongly or even completely suppress the superconducting gap of Nb at the Nb/Py interface [17,20,43]. Hence, no coherence peak of Gilbert damping is expected based on the theoretical study that assumes the completely suppression of the gap at the interface [45]. Besides, the strong suppression of the superconducting gap at the Nb/Py interface is in good agreement of previous tunnel spectroscopy measurements of vanishing superconducting gap at the interface between Nb and Ni [46]. However, for FM insulating GdN, charge carriers from NbN do not penetrate into GdN, thus not weakening the superconductivity at the interface, resulting the survival of the superconducting gap [47,48].

The experimental demonstration of the dynamic spin susceptibility in superconducting thin films via spin pumping could be essential for the field of superconductors. For bulk SCs, dynamic spin susceptibility has been studied from the temperature dependent spin relaxation rate via the nuclear magnetic resonance [43,49,50]. It also provides an avenue for identifying unconventional superconducting paring mechanisms [51], while limited to mostly bulk SCs due to low signal-to-noise ratio [43,49-51]. Spin pumping method has much better signal-to-noise ratio so it could be very sensitive to probe the dynamic spin susceptibility of superconducting thin films. Furthermore, the spin pumping offers a special technique to probe the pair breaking strength, the impurity spin-orbit scattering, and the magnetic proximity effect in the SC/FM junctions, which is also of considerable interest for the field of SC spintronics [1-3].

## IV. CONCLUSION

In conclusion, the spin dynamics of superconducting NbN films are investigated via spin pumping from an adjacent FM insulating layer. A profound coherence peak of the Gilbert damping is observed below $T_C$, which indicates the dynamic spin injection into the superconducting NbN films. Our results paves the way for future investigation of interface impurity spin-orbit scattering, and pair breaking strength in FM/SC junctions, as well as the spin dynamics in the interfacial and two dimensional crystalline SCs. It may also be useful for the search for Majorana fermions in the SC/ferromagnetic insulator heterostructures.

**Acknowledgments**

We acknowledge the fruitful discussion with Yuan Li and Tao Wu. Y.Y., Q.S., W.Y., Y.M., Y.Y., X.C.X., and W.H. acknowledge the financial support from National Basic Research Programs of China (973 program Grant Nos. 2015CB921104 and 2014CB920902) and National Natural Science Foundation of China (NSFC Grant No. 11574006). Y.T., J.P.C, and J.S.M. acknowledge the grants NSF DMR-1700137 and ONR N00014-16-1-2657. Y.T. also acknowledges the JSPS Overseas Research Fellowships and the Fundacion Seneca (Region de Murcia) posdoctoral fellowship (19791/PD/15). W.H. also acknowledges the support by the 1000 Talents Program for Young Scientists of China.

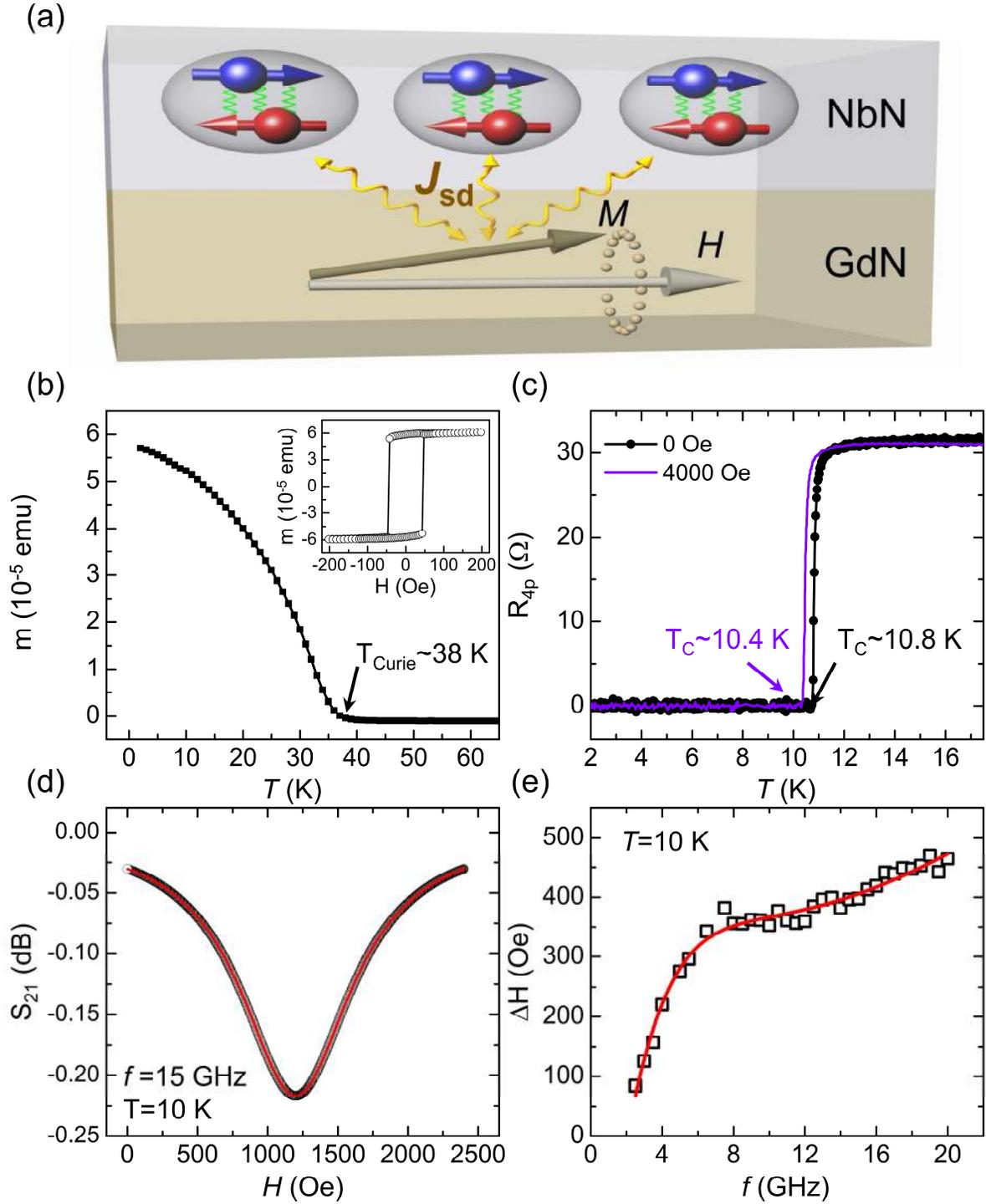

Fig. 1. Spin pumping into the superconducting NbN thin films. (a) Schematic of the interfacial *s*-*d* exchange interaction ($J_{sd}$) between the spins in NbN layer and the rotating magnetization of GdN

layer under the ferromagnetic resonance (FMR) conditions. (b) The magnetic moment as a function of the temperature. Inset: The magnetic hysteresis loop at $T = 5$ K. (c) The four-probe resistance as a function of the temperature with in-plane magnetic field at 0 and 4000 Oe. (d) The typical FMR signal measured at $T = 10$ K and $f = 15$ GHz. The red line indicates the Lorentz fitting curve to obtain the half linewidth based on equation (1). (e) The half linewidth as a function of the microwave frequency at $T = 10$K. The red solid line is the fitting curve based on spin-relaxation model. The results in Fig. 1(b-d) are obtained on the NbN (10)/GdN (5)/NbN (10) sample.

**Figure 2**

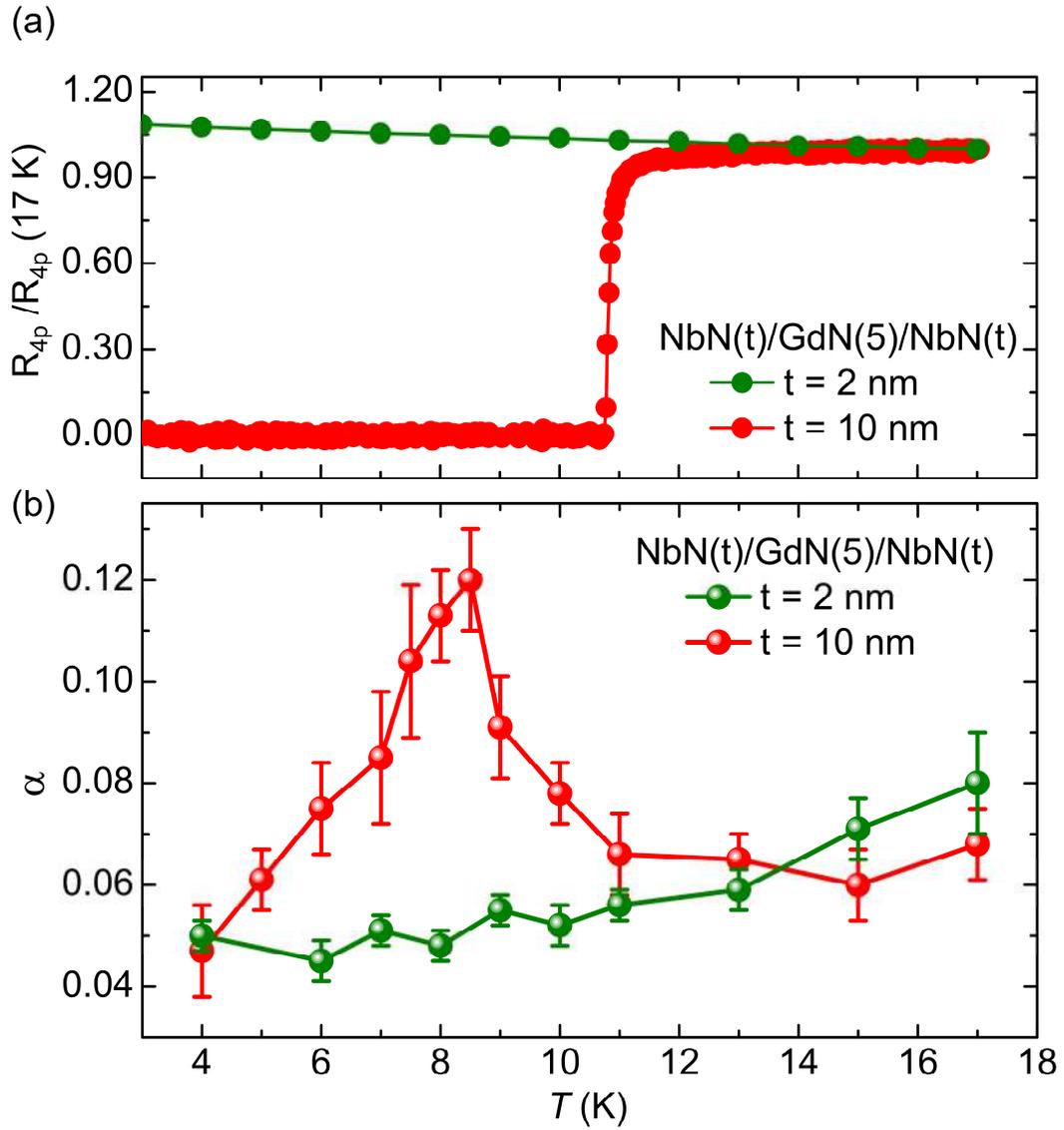

(a)

(b)

Fig. 2. Spin dynamics of the superconducting NbN thin films probed via spin pumping. The normalized four-probe resistance (a) and Gilbert damping (b) as a function of the temperature for the samples of NbN (2)/GdN (5)/NbN (2) and NbN (10)/GdN (5)/NbN (10), respectively.

**Figure 3**

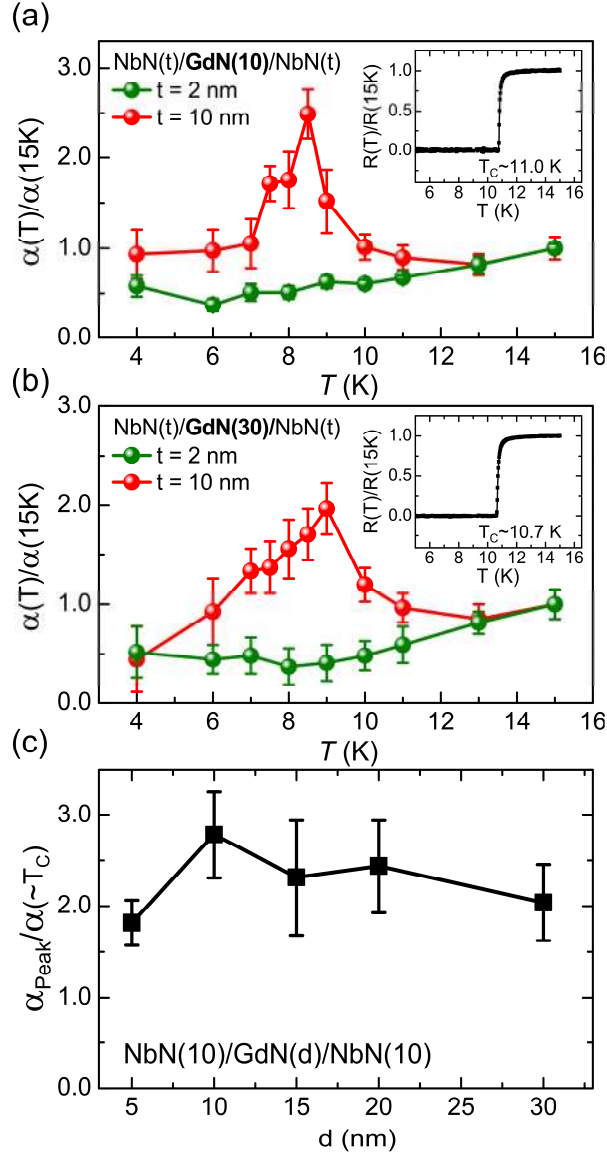

Fig. 3. The GdN thickness effect on spin dynamics of the superconducting NbN thin films. (a-b) The Gilbert damping as a function of the temperature for the samples of NbN (t)/GdN (10)/NbN (t), and NbN (t)/GdN (30)/NbN (t). Insets: The normalized four-probe resistance as a function of the temperature for the samples of NbN (10)/GdN (10)/NbN (10) and NbN (10)/GdN (30)/NbN (10). (c) The ratio of the peak Gilbert damping over the value at $T = \sim T_C$ ($\alpha_{peak} / \alpha(\sim T_C)$) as a function of the GdN layer thickness.

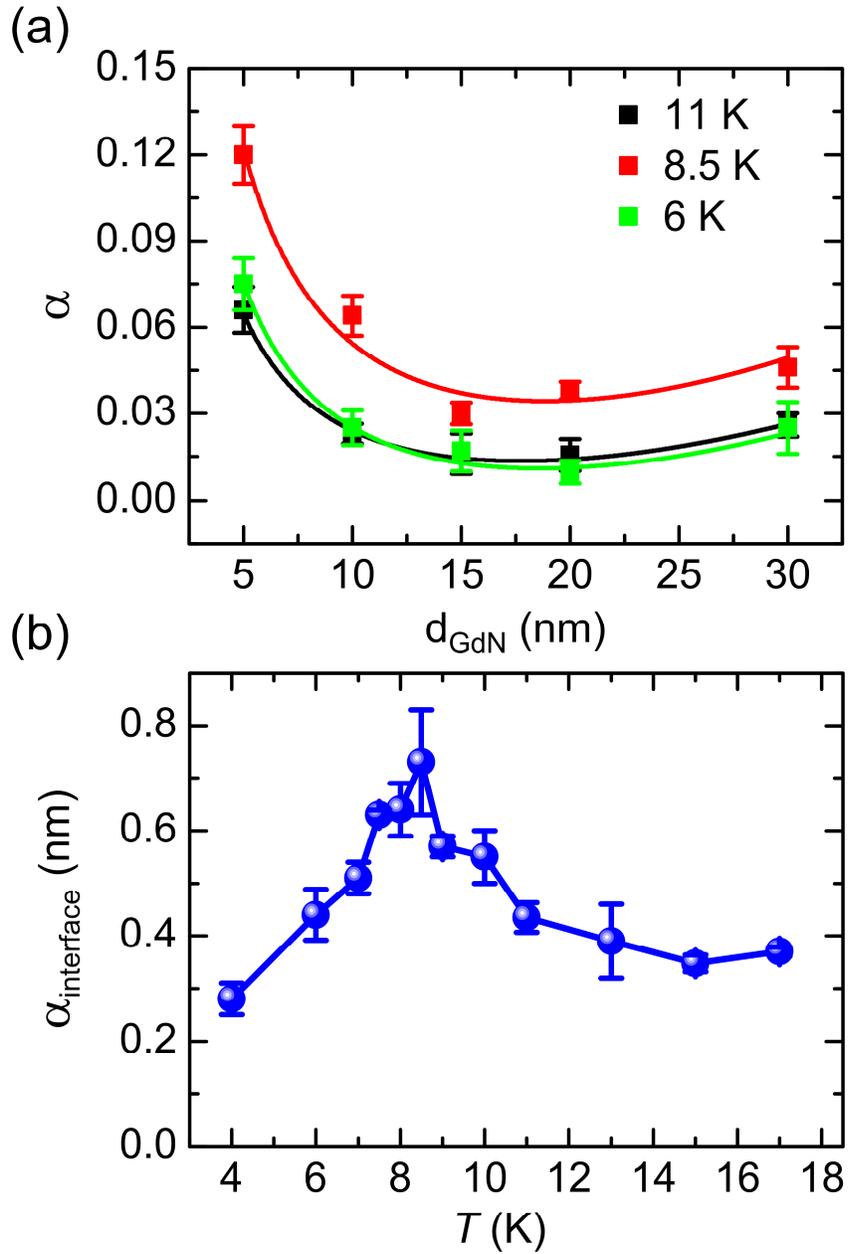

Fig. 4. The interface-induced Gilbert damping at the NbN/GdN interface. (a) The Gilbert damping as a function of the GdN thickness for the samples of NbN (10)/GdN (d)/NbN (10) at $T$ = 11, 8.5, and 6 K, respectively. (b) The interface-induced Gilbert damping as a function of temperature for NbN (10)/GdN (d)/NbN (10).

**Supplementary Materials for:**

**Probe of Spin Dynamics in Superconducting NbN Thin Films via Spin Pumping**

Yunyan Yao[1,2], Qi Song[1,2], Yota Takamura[3,4], Juan Pedro Cascales[3], Wei Yuan[1,2], Yang Ma[1,2], Yu Yun[1,2], X. C. Xie[1,2], Jagadeesh S. Moodera[3,5], and Wei Han[1,2]*

[1]International Center for Quantum Materials, School of Physics, Peking University, Beijing 100871, China.

[2]Collaborative Innovation Center of Quantum Matter, Beijing 100871, China.

[3]Plasma Science and Fusion Center and Francis Bitter Magnet Laboratory, Massachusetts Institute of Technology, Cambridge, MA 02139, USA.

[4]School of Engineering, Tokyo Institute of Technology, Tokyo 152-8550, Japan.

[5]Department of Physics, Massachusetts Institute of Technology, Cambridge, MA 02139, USA

*Correspondence to: weihan@pku.edu.cn (W.H.)

## S1. Determination of Gilbert damping

A nonlinear behavior of half linewidth ($\Delta H$) vs. microwave frequency ($f$) is observed on the NbN (t)/GdN (d)/NbN (t) samples, as shown in Fig. S2(a-c). To our best knowledge, this nonlinear behavior could be attributed to two mechanisms, namely spin-relaxation [38,52], and two-magnon scattering [53,54]. For the spin-relaxation mechanism, $\Delta H$ is related to a temperature-dependent spin-relaxation time constant ($\tau$), and can be expressed by [38]:

$$\Delta H = \Delta H_0 + \frac{4\pi\alpha f}{\gamma} + A\frac{2\pi f\tau}{1+(2\pi f\tau)^2} \qquad (S1)$$

where $\Delta H_0$ is associated with the inhomogeneous properties, $\gamma$ is the gyromagnetic ratio, and $A$ is the spin-relaxation coefficient. While, the relationship of $\Delta H$ and $f$ based on two-magnon scattering mechanism can be expressed by [54,55]:

$$\Delta H = \Gamma\sin^{-1}\sqrt{\frac{\sqrt{(2\pi f)^2+\left(\gamma 2\pi M_{eff}\right)^2}-\gamma 2\pi M_{eff}}{\sqrt{(2\pi f)^2+\left(\gamma 2\pi M_{eff}\right)^2}+\gamma 2\pi M_{eff}}} \qquad (S2)$$

where $\Gamma$ is the two-magnon scattering coefficient, and $M_{eff}$ is the effective magnetization. Clearly, as shown in Fig. S2(a-c), two-magnon scattering mechanism fails to give reasonable fittings (blue lines) to our experimental results (black squares). Whileas, our experimental results agree well with the fitting curves based on the spin-relaxation mechanism (red lines). Besides, it is observed that the nonlinearity of $\Delta H$ vs. $f$ increases as the GdN layer thickness increases, which is also opposite to the two-magnon scattering mechanism. Since two-magnon scattering arises from the defects at the interface, a more conspicuously nonlinear behavior is expected as the FM films become thinner [55-57]. Another feature that $\Delta H$ decreases as $f$ increases (Fig. S2(c)) cannot be explained by the two-magnon scattering mechanism either. Hence, the spin-relaxation mechanism is used to analyze the experimental results of $\Delta H$ vs. $f$, and Gilbert damping could be obtained subsequently.

## S2. Role of the effective magnetization

In previous studies, it has been suggested that the enhancement of the Gilbert damping could be also associated with the change of $M_{eff}$ [35,58]. To rule out the effect of $M_{eff}$, we systemically study the temperature dependence of $M_{eff}$ between the NbN (2)/GdN (d)/NbN (2) and NbN (10)/GdN (d)/NbN (10) samples. The $4\pi M_{eff}$ is based on the Kittel formula [59]:

$$f = (\frac{\gamma}{2\pi})[H_{res}(H_{res} + 4\pi M_{eff})]^{\frac{1}{2}} \tag{S3}$$

where $f$ is the microwave frequency, and $H_{res}$ is the resonance magnetic field. As shown in fig. S4(a), the fitting curve (red line) agrees well with the experimental results. Similar temperature dependences of $4\pi M_{eff}$ for the NbN (2)/GdN (d)/NbN (2) and NbN (10)/GdN (d)/NbN (10) samples are observed (Figs. S4(c-d)). This observation confirms that the observed coherence peak of the Gilbert damping at $T = \sim 8.5$ K is not related to the effective magnetization.

## S3. Determination of interface-induced Gilbert damping

Previous studies have identified the major sources that contribute to the Gilbert damping of the FM films, including bulk damping ($\alpha_B$) and the interface-induced Gilbert damping ($\alpha_S$) that is due to spin pumping and other interface effects [20-22]. As shown in Fig. 4(a) and Fig. S6(a), a slightly increase of the Gilbert damping is observed as the thickness of the GdN layer increases from 20 to 30 nm. This feature could be associated with non-homogeneous field excitation when the width of coplanar wave guide is very small compared to the samples [60,61], and the eddy current loss effect [62]. Both of these effects contribute to the Gilbert damping that is proportional to $d^2$. Hence, the total Gilbert damping can be expressed by:

$$\alpha = \alpha_B + \alpha_S(\frac{1}{d}) + \alpha' d^2 \tag{S4}$$

And the interface-induced Gilbert damping and bulk damping could be obtained subsequently. Based on the fitted results, $\alpha_B$ is significantly small compared to $\alpha_S$. The ~10% error bar of $\alpha_S$ at each temperature makes it hard to obtain the accurate values of $\alpha_B$, which require future studies. Comparing the interface-induced Gilbert damping for NbN (10)/GdN (d)/NbN (10) and NbN (2)/GdN (d)/NbN (2) samples (Fig. 4(b) and Fig. S6(b)), the profound coherence peak is only observed on NbN (10)/GdN (d)/NbN (10) samples at ~ 8.5 K. This observation unambiguously demonstrates that the origin of the coherence peak in the Gilbert damping arises from the interfacial *s-d* exchange interaction between the magnetic moments of GdN and the spins of the quasiparticles of NbN in its superconducting state.

## S4. The impact of spin pumping experiments on the $T_C$

Beyond the magnetic field effect on the superconducting films (Fig. 1(c)), the spin pumping experiments with various microwave excitations do not affect the superconducting films either. As shown in Figs. S8(a) and S8(b), the $T_C$ exhibit a small variation probed on the typical sample NbN (10)/GdN (5)/NbN (10).

Figure S1

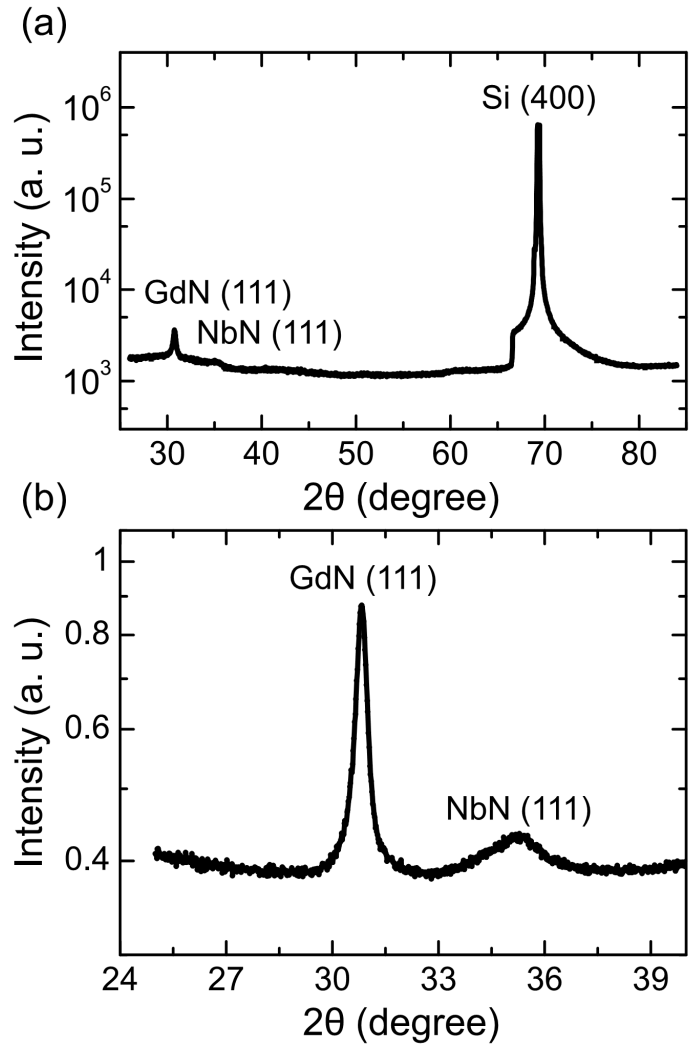

Figure S1. Crystalline structure of the (111)-textured GdN and NbN films. (a-b), XRD results measured on the NbN (10)/GdN (50)/NbN (10) sample. Two main observable peaks correspond to GdN (111) and NbN (111).

Figure S2

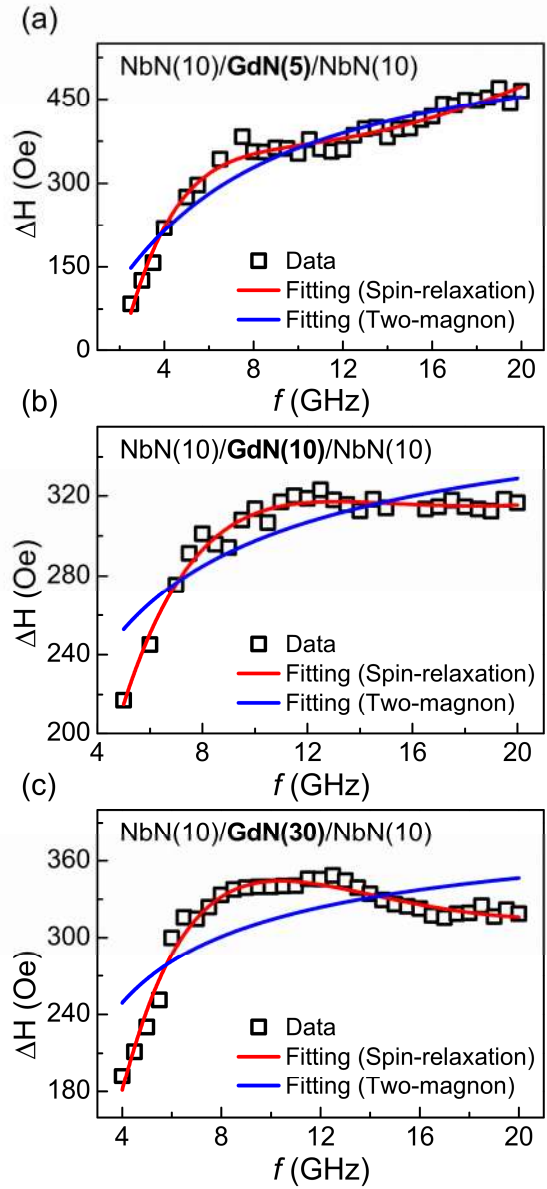

Figure S2. Comparison of the spin-relaxation and two-magnon scattering mechanisms. (a-c) The experimental results of ΔH vs. *f* at *T* = 10 K and the fitting curves (red/green lines for the spin relaxation/two-magnon scattering) on NbN (10)/GdN (d)/NbN (10) samples with d = 5, 10, and 30 nm respectively.

Figure S3

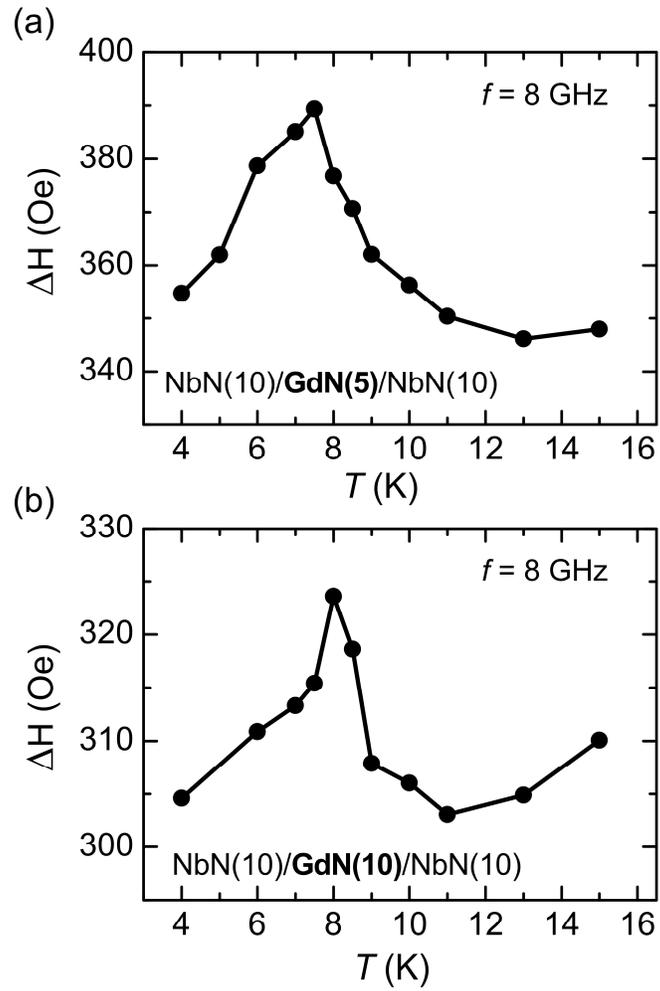

Figure S3. Half linewidth as a function of temperature for NbN (10)/GdN (5)/NbN (10) (a) and NbN (10)/GdN (10)/NbN (10) (b) samples.

Figure S4

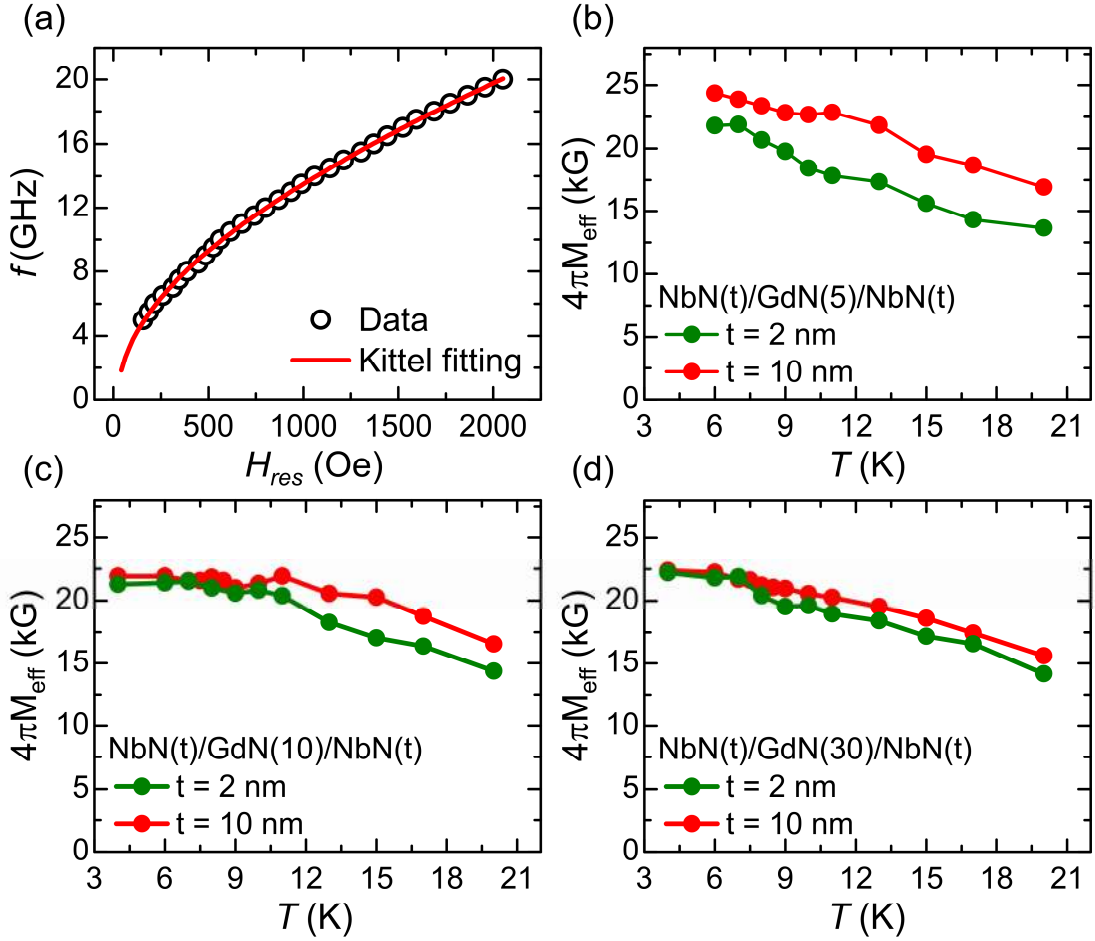

Figure S4. Characterization of effective magnetization. (a) The determination of $M_{eff}$ via Kittel formula (red line) from the experimental results of $f$ vs. $H_{res}$. (b-d), $4\pi M_{eff}$ as a function of temperature for the NbN (2)/GdN (d)/NbN (2) (green circles) and NbN (10)/GdN (d)/NbN (10) samples (red circles) with d = 5 nm (b), 10 nm (c), and 30 nm (d), respectively.

Figure S5

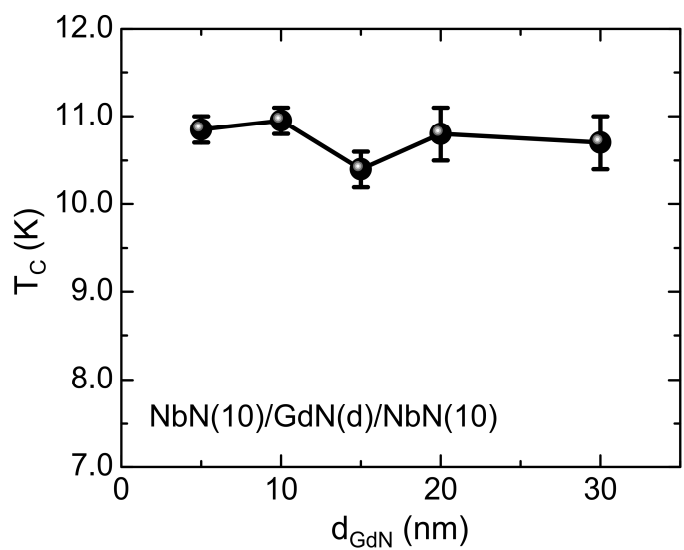

Figure S5. T$_C$ of NbN as a function of the GdN thickness. These results are obtained on the NbN (10)/GdN (d)/NbN (10) samples.

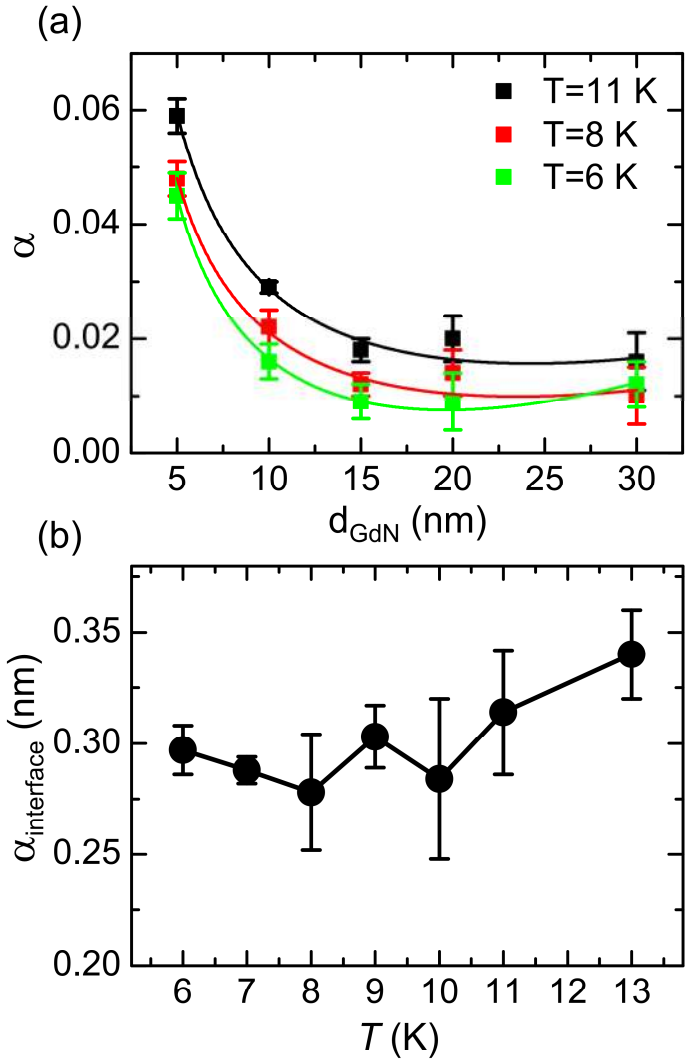

Figure S6. Interface-induced Gilbert damping in NbN (2)/GdN (d)/NbN (2). (a) The Gilbert damping constant as a function of the GdN thickness at $T$ = 11, 8, and 6 K, respectively. Solid lines are the fitting curves based on the equation (S4). (b) The interface-induced Gilbert damping as a function of temperature.

Figure S7

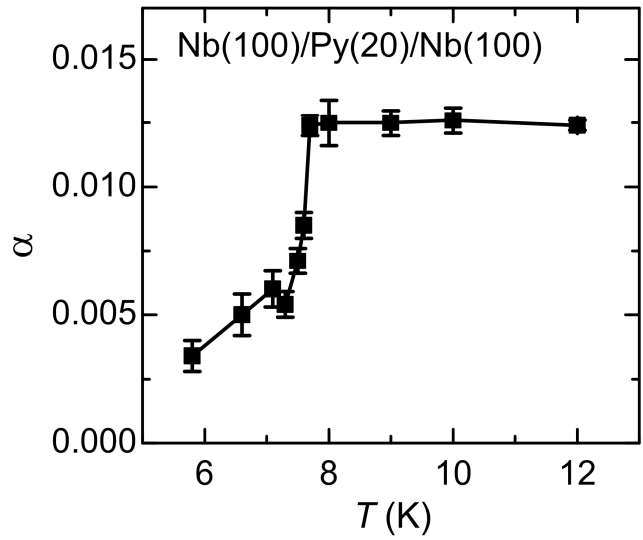

Figure S7. Temperature dependence of Gilbert damping of Py in the Nb (100)/Py (20)/Nb (100) sample.

Figure S8

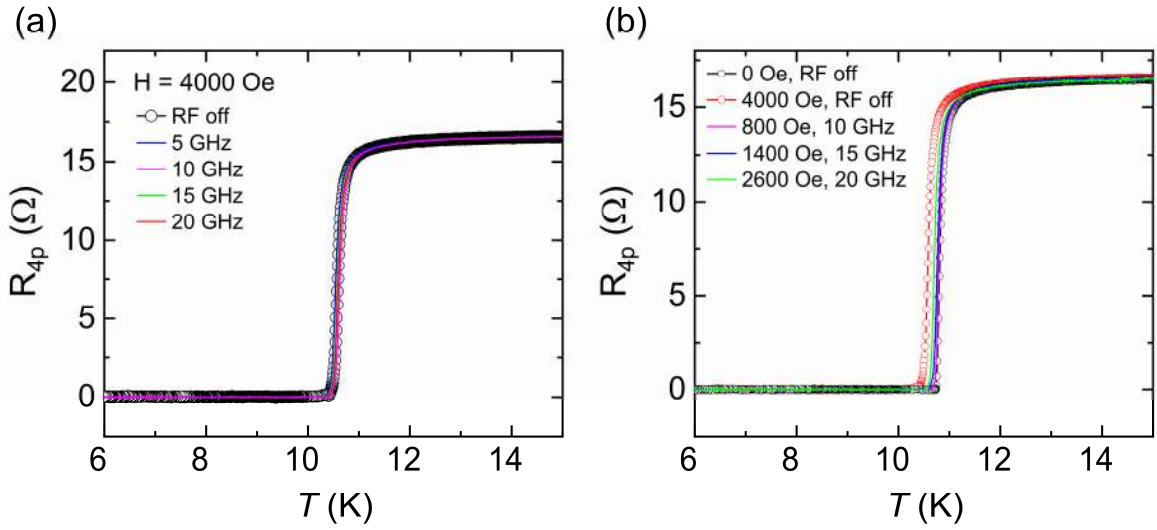

Figure S8. The impact of spin pumping experiments on the T$_C$ on the typical NbN (10)/GdN (5)/NbN (10) sample. (a) The four-probe resistance vs. temperature under 4000 Oe with various microwave excitation frequencies. (b) The four-probe resistance vs. temperature around the FMR resonance conditions of GdN.